\documentclass[11 pt]{elsarticle}
\usepackage{amssymb, amsmath}
\usepackage{algorithm}
\usepackage{algorithmic}
\usepackage{verbatim}   
\usepackage[margin = 0.75 in]{geometry}
\usepackage[pdftex, pdfstartview={FitH}, pdfnewwindow=true, colorlinks=true, citecolor=blue, filecolor=blue, linkcolor=blue, urlcolor=blue, pdfpagemode=UseNone]{hyperref}

\usepackage{listings}
\def\figheight{7 cm}
\newcommand{\Cbb}{{\mathbb C}}
\newcommand{\Ibb}{{\mathbb I}}

\newcommand{\muhat}{\ensuremath{\widehat \mu} }
\newcommand{\Gdag}{\ensuremath{G^{\dag}} }
\newcommand{\al}{\ensuremath{\alpha} }
\newcommand{\aldot}{\ensuremath{\dot\al} }
\newcommand{\be}{\ensuremath{\beta} }
\newcommand{\de}{\ensuremath{\delta} }
\newcommand{\De}{\ensuremath{\Delta} }
\newcommand{\eps}{\ensuremath{\epsilon} }
\newcommand{\la}{\ensuremath{\lambda} }
\newcommand{\ka}{\ensuremath{\kappa} }

\newcommand{\Phidag}{\ensuremath{\Phi^{\dag}} }

\newcommand{\si}{\ensuremath{\sigma} }
\newcommand{\cA}{\ensuremath{\mathcal A} }

\newcommand{\cD}{\ensuremath{\mathcal D} }
\newcommand{\cDbar}{\ensuremath{\overline{\mathcal D}} }
\newcommand{\cDdag}{\ensuremath{\mathcal D^{\dag}} }
\newcommand{\cF}{\ensuremath{\mathcal F} }
\newcommand{\cFbar}{\ensuremath{\overline{\mathcal F}} }
\newcommand{\cO}{\ensuremath{\mathcal O} }
\newcommand{\cP}{\ensuremath{\mathcal P} }
\newcommand{\cN}{\ensuremath{\mathcal N} }
\newcommand{\cQ}{\ensuremath{\mathcal Q} }
\newcommand{\cU}{\ensuremath{\mathcal U} }
\newcommand{\cV}{\ensuremath{\mathcal V} }
\newcommand{\cUbar}{\ensuremath{\overline{\mathcal U}} }
\newcommand{\cZ}{\ensuremath{\mathcal Z} }
\newcommand{\lra}{\ensuremath{\longrightarrow} }
\newcommand{\Lra}{\ensuremath{\Longrightarrow} }
\newcommand{\nn}{\nonumber}
\newcommand{\X}{\ensuremath{\!\times\!} }
\newcommand{\KD}{K\"ahler--Dirac }
\newcommand{\glN}{\ensuremath{\mathfrak{gl}(N, \Cbb)} }
\newcommand{\slN}{\ensuremath{\mathfrak{sl}(N, \Cbb)} }
\newcommand{\sixt}{\ensuremath{16^3\X32} }
\newcommand{\SLAT}{\texttt{SUSY~LATTICE}}
\newcommand{\NCOL}{\texttt{NCOL}}
\newcommand{\DIMF}{\texttt{DIMF}}
\newcommand{\ds}{\texttt{ds}}
\newcommand{\bc}{\texttt{bc}}
\newcommand{\pizero}{\texttt{pi0}}
\newcommand{\vev}[1]{\ensuremath{\left\langle #1 \right\rangle} }
\newcommand{\deriv}[2]{\ensuremath{\frac{\de #1}{\de #2}} }
\newcommand{\diag}{\ensuremath{\mbox{diag}} }
\newcommand{\pf}[0]{\ensuremath{\mbox{pf}\,} }
\newcommand{\Tr}[1]{\ensuremath{\mbox{Tr}\left[ #1 \right]} }
\newcommand{\eq}[1]{Eq.~\ref{#1}}
\newcommand{\fig}[1]{Fig.~\ref{#1}}
\newcommand{\refcite}[1]{Ref.~\cite{#1}}
\newcommand{\secref}[1]{Section~\ref{#1}}

\begin{document}
\begin{frontmatter}

\title{Parallel software for lattice $\cN = 4$ supersymmetric Yang--Mills theory}
\author[DS]{David Schaich}\ead{dschaich@syr.edu}
\author[TD]{Thomas DeGrand}\ead{thomas.degrand@colorado.edu}
\address[DS]{Department of Physics, Syracuse University, Syracuse, New York 13244, United States}
\address[TD]{Department of Physics, University of Colorado, Boulder, Colorado 80309, United States}

\begin{abstract} 
  We present new parallel software, \SLAT, for lattice studies of four-dimensional $\cN = 4$ supersymmetric Yang--Mills theory with gauge group SU($N$).
  The lattice action is constructed to exactly preserve a single supersymmetry charge at non-zero lattice spacing, up to additional potential terms included to stabilize numerical simulations.
  The software evolved from the MILC code for lattice QCD, and retains a similar large-scale framework despite the different target theory.
  Many routines are adapted from an existing serial code~\cite{Catterall:2011cea}, which \SLAT\ supersedes.
  This paper provides an overview of the new parallel software, summarizing the lattice system, describing the applications that are currently provided and explaining their basic workflow for non-experts in lattice gauge theory.
  We discuss the parallel performance of the code, and highlight some notable aspects of the documentation for those interested in contributing to its future development. \\
\end{abstract}

\end{frontmatter}

\section{\label{sec:intro}Introduction: Supersymmetry on the lattice} 
For decades, lattice field theory has played important roles in many areas of physics, from statistical and condensed-matter physics to high-energy particle physics.
Within the domain of particle physics, the formulation of gauge theories on a space-time lattice provides a non-perturbative definition of such systems, with large-scale numerical computations making crucial contributions to the phenomenology of quantum chromodynamics (QCD) and the search for new physics beyond the standard model (BSM).\footnote{The proceedings of the annual International Symposium on Lattice Field Theory provide comprehensive reviews and may be found at \url{http://pos.sissa.it/cgi-bin/reader/family.cgi?code=lattice}.}

An area where progress has been more difficult is the lattice formulation of supersymmetric theories~\cite{Catterall:2009it}.
This is unfortunate, because supersymmetry is an extremely important theoretical tool, and a common ingredient in BSM physics.
The simplifications resulting from supersymmetry have aided analytic investigations of many fascinating non-perturbative phenomena such as confinement, dynamical symmetry breaking, and dualities between pairs of supersymmetric gauge theories as well as between gauge and gravity theories~\cite{Terning:2006bq}.
Complementary studies of these issues through first-principles lattice calculations, in addition to potential phenomenological applications to supersymmetric extensions of the standard model, motivate continuing research into supersymmetric lattice gauge theory.

Briefly stated, the essential obstacle to lattice supersymmetry is the fact that supersymmetry transformations extend the algebra of space-time translations and rotations.
The anticommutators of supersymmetry generators produce infinitesimal translations, $\left\{Q_{\al}, Q_{\aldot}^{\dag}\right\} \propto \si_{\al\aldot}\cdot P$, which do not exist in discrete space-time.
This problem cannot be evaded by defining a ``discrete supersymmetry'' based on finite lattice translations, because lattice finite-difference operators do not satisfy the Leibniz rule~\cite{Dondi:1976tx, Kato:2008sp}.
Naive lattice discretization thus breaks supersymmetry at the classical level, leading to the generation of supersymmetry-violating operators whose couplings must be fine-tuned to recover the supersymmetric theory of interest.
Since there are typically many such operators, such fine-tuning is generally impractical.\footnote{One notable exception is $\cN = 1$ supersymmetric Yang--Mills theory in four dimensions, where only the gaugino mass needs to be fine-tuned~\cite{Bergner:2013nwa}, or controlled by working with Ginsparg--Wilson lattice fermions~\cite{Giedt:2008xm, Endres:2009yp, Kim:2011fw}.}

For certain gauge theories with extended supersymmetry, however, significant progress was made in the past decade through the construction of lattice formulations in which a subset of the supersymmetry algebra is exactly preserved at non-zero lattice spacing.\footnote{There are, of course, other approaches that lie beyond the scope of this article~\cite{Ishii:2008ib, Ishiki:2008te, Ishiki:2009sg, Hanada:2010kt, Honda:2011qk, Honda:2013nfa, Hanada:2013rga}.}
The historical development and essential features of these constructions are thoroughly reviewed in \refcite{Catterall:2009it}.
The systems for which this approach is possible include $\cN = (2, 2)$, $(4, 4)$ and $(8, 8)$ supersymmetric Yang--Mills (SYM) theories in two dimensions, $\cN = 4$ SYM in three dimensions, and the especially interesting case of $\cN = 4$ SYM in four dimensions.
For all of these theories, the method of topological twisting provides a change of variables under which the supercharges form \KD multiplets of antisymmetric $p$-index tensors $\left(\cQ, \cQ_{\mu}, \cQ_{\mu\nu}, \cdots\right)$ with $p = 0, 1, \cdots, d$ in $d$ dimensions.
The nilpotent 0-form supercharges \cQ anticommute to form a sub-algebra that can be exactly preserved on the lattice~\cite{Kaplan:2005ta, Unsal:2006qp}.
This procedure does not depend on the gauge group, and in this article we consider SU($N$) gauge theories with relatively small numbers of colors $N = 2$, 3 and 4.

More recently, these constructions have been used as the basis for practical numerical lattice calculations, which have passed several non-trivial consistency checks and begun to make novel contributions to our physical understanding~\cite{Catterall:2011aa, Catterall:2012yq, Catterall:2014vka, Catterall:2014mha}.
Especially in the most interesting case of four-dimensional $\cN = 4$ SYM, these lattice studies rapidly become computationally expensive, a challenge familiar from lattice QCD.
To address this challenge, we have developed new parallel software for lattice $\cN = 4$ SYM, \SLAT, which we introduce in this article.
Our starting point was the highly-optimized MIMD Lattice Computation (MILC) software for lattice QCD~\cite{MILC:manual}, although the significantly different target theory required substantial changes throughout the code.
Many central routines are adapted from the serial code presented in \refcite{Catterall:2011cea}, which this software supersedes, and we have also added some new capabilities that extend the scope of the supported computations.

In this work we present an overview of \SLAT, aiming to provide useful information both for particle theorists with limited Monte Carlo experience as well as for lattice gauge theorists more familiar with QCD-like systems.
We begin in the next section by briefly reviewing the lattice formulation of $\cN = 4$ SYM, and the basic workflow of lattice investigations.
Summarizing discussions recently published in \refcite{Catterall:2014vka}, we describe the degrees of freedom of the discretized system, the $A_4^*$ lattice structure, and the lattice action implemented by \SLAT, including potential terms that softly break the otherwise exactly preserved 0-form supersymmetry $\cQ$.
In \secref{sec:RHMC} we explain the rational hybrid Monte Carlo (RHMC) algorithm for importance sampling of gauge field configurations and introduce a few important measurements, in particular the phase of the complex Pfaffian of the fermion operator.
For users interested in obtaining and running \SLAT\ without modification, in \secref{sec:run} we provide a quick-start guide including a high-level overview of the available measurements, along with some discussion of the scaling and parallel performance of the code.
Finally, in \secref{sec:code} we highlight some notable aspects of the documentation for those who may be interested in extending \SLAT\ with additional measurements or features.
We conclude with a discussion of some potential directions for future development, and invite all who wish to contribute to do so through the publicly accessible version control repository~\cite{GitHub}.

\section{\label{sec:physics}The discretized theory, and methodology of lattice calculations} 
In this section we first write down the lattice formulation of $\cN = 4$ SYM in four dimensions, and then review the essential ingredients of lattice gauge theory calculations.
We omit the details of the relevant topological twisting of continuum $\cN = 4$ SYM~\cite{Marcus:1995mq, Kapustin:2006pk}, which is discussed in more detail in Refs.~\cite{Catterall:2009it, Catterall:2014vka}.
Instead, in \secref{sec:lattice} we simply introduce the degrees of freedom of the lattice theory, describe the $A_4^*$ lattice structure, and write down the lattice action.
We explain the various terms in the lattice action, emphasizing their invariance under lattice gauge transformations and (except for certain potential terms) preservation of the 0-form twisted supersymmetry $\cQ$.

Given the lattice action $S$, we can carry out numerical computations of operator expectation values
\begin{align}
  \label{eq:path_int}
  \vev{\cO} & = \frac{1}{\cZ} \int[dX] \,\cO\, e^{-S[X]} &
  \cZ & = \int[dX] e^{-S[X]},
\end{align}
where $X$ is a placeholder for the lattice fields to be defined below.
\secref{sec:RHMC} summarizes how the discretized path integral is stochastically evaluated through importance-sampling Monte Carlo.
In broadest terms, we use the rational hybrid Monte Carlo (RHMC) algorithm to obtain a sequence of gauge field configurations with the appropriate probability distribution.
We then measure observables of interest on a set of $n$ such configurations, so that $\vev{\cO} = \frac{1}{n}\sum_i \cO_i$, with statistical uncertainties from the finite sample size and systematic uncertainties from working in a finite discretized space-time.
In \secref{sec:RHMC} we will discuss some important observables, including the low-lying eigenvalues and Pfaffian of the fermion operator.

\subsection{\label{sec:lattice}Lattice variables, lattice structure and lattice action} 
Continuum $\cN = 4$ SYM is a theory of a gauge field, four Majorana fermion fields and six scalar fields, all of which transform in the adjoint representation of the SU($N$) gauge group, and are related to each other by 16 fermionic supersymmetry charges.
Working in euclidean space-time, the twisting procedure regroups the ten bosons into a five-component complexified gauge field $\cA_a$ with $a = 0, \cdots, 4$, while the 16 fermions are assigned to the multiplet $\left(\eta, \psi_a, \chi_{ab}\right)$, with $\chi_{ab}$ antisymmetric.
The supersymmetry charges are similarly converted to $\left(\cQ, \cQ_a, \cQ_{ab}\right)$.

At this point we can move onto the lattice~\cite{Kaplan:2005ta, Unsal:2006qp, Catterall:2007kn, Damgaard:2008pa, Catterall:2014vka}, by defining complexified gauge links $\cU_a(n)$ that are elements of the algebra $\glN$,
\begin{equation}
  \cU_a(n) = \sum_{C = 0}^{N^2 - 1} T^C \cU_a^C(n),
\end{equation}
with complex coefficients $\cU_a^C(n)$. 
The $T^C$ are antihermitian generators of $\mathfrak u(N)$, with normalization $\Tr{T^A T^B} = -\de^{AB}$.
The fermion fields $\eta(n)$, $\psi_a(n)$ and $\chi_{ab}(n)$ are defined in the same way, as required by supersymmetry.
As a consequence, the fields transform in the adjoint representation of U($N$),
\begin{align}
  \cU_a(n) & \to G(n) \cU_a(n) \Gdag(n + \muhat_a)        & \eta(n) & \to G(n) \eta(n) \Gdag(n)                                       \cr
  \cUbar_a(n) & \to G(n + \muhat_a) \cUbar_a(n) \Gdag(n)  & \psi_a(n) & \to G(n) \psi_a(n) \Gdag(n + \muhat_a) \label{eq:gaugeTrans}  \\
                                                        & & \chi_{ab}(n) & \to G(n + \muhat_a + \muhat_b) \chi_{ab}(n) \Gdag(n),      \nn
\end{align}
for lattice gauge transformation $G \in \mbox{U}(N)$.
Speaking informally, we will say that the link $\cU_a(n)$ points ``from'' site $n + \muhat_a$ (on the right) ``to'' site $n$ (on the left), and similarly for the other fields.

Because we have five gauge links going out from each lattice site, we cannot discretize the theory on the familiar hypercubic lattice.
Instead we must use the $A_4^*$ lattice, the four-dimensional lattice where each site has ten symmetric nearest neighbors connected by five linearly dependent basis vectors.
(The $A_d^*$ lattice in $d$ dimensions has coordination number $2(d + 1)$; the most familiar example is the triangular lattice $A_2^*$, while $A_3^*$ can be deformed into the body-centered cubic lattice.)
In \SLAT\ we employ an extremely convenient representation of the $A_4^*$ lattice, which simply adds a body-diagonal link $\muhat_4 = (-1, -1, -1, -1)$ to the usual four hypercubic basis vectors $\muhat_{\nu}$.
Converting from this abstract hypercubic basis to physical space-time requires some care, as discussed in \refcite{Catterall:2014vka}.

Now we can write down the full lattice action (repeated indices summed except where noted),
\begin{align}
  S           & = S_{exact} + S_{closed} + S_{stab}                                                                                                                                                                 \label{eq:lattice_action} \\
  S_{exact}   & = \frac{N}{2\la} \sum_n a^4 \ \Tr{-\cFbar_{ab}(n)\cF_{ab}(n) + \frac{c_2}{2}\left(\cDbar_a^{(-)}\cU_a(n)\right)^2 - \chi_{ab}(n) \cD_{[a}^{(+)}\psi_{b]}^{\ }(n) - \eta(n) \cDbar_a^{(-)}\psi_a(n)} \label{eq:Sexact}         \\
  S_{closed}  & = -\frac{N}{8\la} \sum_n a^4 \ \Tr{\epsilon_{abcde}\ \chi_{de}(n + \muhat_a + \muhat_b + \muhat_c) \cDbar_c^{(-)} \chi_{ab}(n)}                                                         \label{eq:Sclosed}        \\
  S_{stab}    & = \frac{N}{2\la}\mu^2 \sum_{n,\ c} a^4 \left(\frac{1}{N}\Tr{\cU_c(n) \cUbar_c(n)} - 1\right)^2 + \ka \sum_{\cP} a^4 |\det \cP - 1|^2.                                                               \label{eq:Sstab}
\end{align}
These require some further definitions.
The field strengths $\cF_{ab}$ and $\cFbar_{ab}$ in the first term are
\begin{equation}
  \begin{split}
    \cF_{ab}(n) & = \cU_a(n) \cU_b(n + \muhat_a) - \cU_b(n) \cU_a(n + \muhat_b) = \cD_a^{(+)} \cU_b(n) \\
    \cFbar_{ab}(n) & = \cUbar_a(n + \muhat_b) \cUbar_b(n) - \cUbar_b(n + \muhat_a) \cUbar_a(n) = \cDbar_a^{(+)} \cUbar_b(n),
  \end{split}
\end{equation}
while the forward and backward finite-difference operators are~\cite{Catterall:2007kn, Damgaard:2008pa}
\begin{align}
  \cD_a^{(+)} f_b(n) & = \cU_a(n) f_b(n + \muhat_a) - f_b(n) \cU_a(n + \muhat_b)                                          \cr
  \cDbar_a^{(-)} f_a(n) & = f_a(n) \cUbar_a(n) - \cUbar_a(n - \muhat_a) f_a(n - \muhat_a)                                 \\
  \cDbar_a^{(+)} f_b(n) & = \cUbar_a(n + \muhat_b) f_b(n) - f_b(n + \muhat_a) \cUbar_a(n)                                 \cr
  \cDbar_c^{(-)} f_{ab}(n) & = f_{ab}(n + \muhat_c) \cUbar_c(n) - \cUbar_c(n + \muhat_a + \muhat_b)f_{ab}(n).  \nn
\end{align}
In the final term of \eq{eq:Sstab}, \cP is the oriented product of links around a fundamental plaquette of the lattice.
For the remainder of this article we will work in lattice units where we set the lattice spacing $a = 1$.

With these definitions it is straightforward to confirm that the lattice action $S$ is gauge invariant: all terms in \eq{eq:lattice_action} form closed loops, and $\Gdag G = G\Gdag = 1$ for gauge transformation $G \in \mbox{U}(N)$.
With the exception of $S_{stab}$ (discussed below), $S$ also preserves the 0-form supersymmetry $\cQ$, so that $\cQ S = 0$ when $\mu = 0$ and $\ka = 0$.
Under the action of \cQ the fields transform as
\begin{align}
  & \cQ\; \cU_a(n) = \psi_a(n)            & & \cQ\; \psi_a(n) = 0                   \cr
  & \cQ\; \chi_{ab}(n) = -\cFbar_{ab}(n)  & & \cQ\; \cUbar_a(n) = 0 \label{eq:susy} \\
  & \cQ\; \eta(n) = d(n)                  & & \cQ\; d(n) = 0                        \nn
\end{align}
with $d$ a bosonic auxiliary field that maintains off-shell supersymmetry.
In \eq{eq:lattice_action} we replaced $d$ by its equation of motion $d(n) \to \cDbar_a^{(-)} \cU_a(n)$.
From \eq{eq:susy} it is clear that $\cQ^2 = 0$ on every field, and it is not hard to see that \cQ acting on $S_{exact}$ vanishes.
The action of \cQ on $S_{closed}$ also vanishes due to a lattice Bianchi identity, $\eps_{abcde} \cDbar_c \cFbar_{de} = 0$.

The other fifteen supersymmetries $\cQ_a$ and $\cQ_{ab}$ are broken on the lattice, and must be restored in the continuum limit to recover full $\cN = 4$ SYM.
This requirement is analogous to the restoration of SO(4) euclidean Lorentz symmetry in the continuum limit of lattice QCD calculations, where only a discrete subgroup of SO(4) is preserved at non-zero lattice spacing.

Finally, we define the parameters in the lattice action, \eq{eq:lattice_action}.
The overall factor outside most of the terms in $S$ involves the number of colors $N$ and the 't~Hooft coupling $\la = g^2 N$, where $g$ is the gauge coupling.
The coefficient $c_2$ in \eq{eq:Sexact} takes the value $c_2 = 1$ classically, but may be shifted by quantum effects, potentially requiring fine-tuning to recover $\cN = 4$ SYM in the continuum limit.
Preliminary investigations suggest that the necessary fine-tuning may be negligible in practice~\cite{Catterall:2014mha}, and we typically fix $c_2 = 1$.

$S_{stab}$ with its two coefficients $\mu^2$ and \ka is used to stabilize numerical computations.
The ``bosonic mass'' $\mu$ regulates flat directions, lifting a bosonic zero mode present due to the periodic boundary conditions (BCs) in all four directions.
We lift the corresponding fermionic zero mode by imposing antiperiodic (thermal) temporal BCs for the fermions.
Alternately, a fermion mass term
\begin{equation}
  \label{eq:fmass}
  m \sum_n a^4 \left\{\Tr{\cDbar_b^{(-)}\cU_b(n)}\Tr{\cU_a(n) \cUbar_a(n)} - \Tr{\eta(n)}\Tr{\psi_a(n) \cUbar_a(n)}\right\}
\end{equation}
could be added to the action, but this is not currently implemented in \SLAT.
Finally, when $\ka > 0$ the plaquette determinant term projects the product of links around each plaquette \cP from \glN to $\slN$, effectively reducing the gauge group from U($N$) to the target SU($N$).
Specifically, $\ka \geq 0.5$ forbids monopole condensation in the U(1) sector, protecting the lattice system from these unphysical strong-coupling lattice artifacts~\cite{Catterall:2014vka}.

Non-zero $\mu$ and \ka softly break the 0-form supersymmetry $\cQ$, which is otherwise exactly preserved even at non-zero lattice spacing: $\lim_{(\mu, \ka) \to (0, 0)} \cQ S = 0$.
(The fermion mass term in \eq{eq:fmass} is constructed in a $\cQ$-invariant way, but would break a global shift symmetry under $\eta \to \eta + c\Ibb$ with $c$ a constant Grassmann parameter.) 
The bosonic mass $\mu$ must be tuned to zero in the continuum limit to recover $\cN = 4$ SYM.
It may not be necessary to tune \ka to zero, since its $\cQ$-breaking effects are confined to the U(1) sector that decouples in the continuum limit where $\mbox{U}(N) = \mbox{SU}(N) \otimes \mbox{U}(1)$.
In the next subsection we will discuss ways to monitor the effects of non-zero $\mu$ and \ka in numerical calculations, along with ways to check the restoration of the broken supersymmetries $\cQ_a$ and $\cQ_{ab}$.

\subsection{\label{sec:RHMC}Rational hybrid Monte Carlo algorithm and important observables} 
Given the lattice formulation of $\cN = 4$ SYM written above, we wish to compute operator expectation values through importance sampling Monte Carlo evaluation of the path integral in \eq{eq:path_int}.
Let us divide the lattice action into its bosonic and fermionic parts,
\begin{equation}
  S[\cU, \cUbar, \Psi] = S_B[\cU, \cUbar] + \Psi^T \cD[\cU, \cUbar] \Psi
\end{equation}
where $\Psi \equiv \left(\eta, \psi_a, \chi_{ab}\right)^T$ is the \KD fermion field and $\cD[\cU, \cUbar]$ is the (antisymmetric) fermion operator.
\eq{eq:path_int} then becomes
\begin{equation}
  \vev{\cO} = \frac{1}{\cZ} \int[d\cU] [d\cUbar] [d\Psi] \,\cO\, e^{-S[\cU, \cUbar, \Psi]} = \frac{1}{\cZ} \int[d\cU] [d\cUbar]  \,\cO\, e^{-S_B[\cU, \cUbar]}\ \pf \cD[\cU, \cUbar],
\end{equation}
where the Pfaffian $\pf \cD[\cU, \cUbar]$ results from Gaussian integration over $\Psi$.

Because we wish to employ $[e^{-S_B}\, \pf \cD]$ as a Boltzmann weight, both the bosonic action and Pfaffian should be real and positive.
This condition is satisfied for $S_B$.
Unfortunately, for a given gauge field configuration the Pfaffian is generically complex, $\pf \cD = |\pf \cD|e^{i\al}$.
We proceed by considering the phase-quenched path integral
\begin{align}
  \vev{\cO}_{pq} & = \frac{1}{\cZ_{pq}} \int[d\cU] [d\cUbar] \,\cO\, e^{-S_B[\cU, \cUbar]}\ |\pf \cD[\cU, \cUbar]| &
  \cZ_{pq} & = \int[d\cU] [d\cUbar] e^{-S_B[\cU, \cUbar]}\ |\pf \cD[\cU, \cUbar]|.
\end{align}
So long as $\vev{e^{i\al}}_{pq}$ is statistically non-zero, the true expectation values can be reconstructed via phase reweighting,
\begin{equation}
  \vev{\cO} = \frac{\vev{\cO e^{i\al}}_{pq}}{\vev{e^{i\al}}_{pq}}.
\end{equation}
The phase of the Pfaffian is therefore a crucial observable to monitor in phase-quenched calculations, and we will discuss its measurement in detail in \secref{sec:Pfaffian}.
In practice, we find that the Pfaffian of lattice $\cN = 4$ SYM is very nearly real and positive, so that $\vev{\cO}_{pq} \approx \vev{\cO}$ and we do not suffer from an insurmountable sign problem~\cite{Catterall:2014vka}.
Accordingly, we will omit the $_{pq}$ subscripts in the remainder of this article.

Explicitly evaluating the Pfaffian is far too computationally expensive to be a part of our configuration generation algorithm.
The standard approach we use instead employs a set of bosonic pseudofermions $\Phi$, of the same dimensionality as $\Psi$, which are introduced through the mathematical identity
\begin{equation}
  \label{eq:pf}
  |\pf \cD| = |\det \cD|^{1 / 2} = \left(\det[\cDdag \cD]\right)^{1 / 4} \propto \int[d\Phidag][d\Phi] \exp\left[-\Phidag \left(\cDdag \cD\right)^{-1 / 4} \Phi\right].
\end{equation}
The negative fractional power is achieved by the use of the rational hybrid Monte Carlo (RHMC) algorithm to generate an appropriate distribution of gauge configurations~\cite{Clark:2006fx}.

There are three main steps to the RHMC algorithm.
First the pseudofermion field is set to $\Phi(n) = \left(\cDdag \cD\right)^{1 / 8} g(n)$ at each lattice site, with $g(n)$ being $16N^2$-component random vectors drawn from a Gaussian distribution.
We also define random Gaussian $\Pi_a(n) \in \glN$ to serve as fictitious momenta conjugate to the gauge fields $\cU_a(n)$, leading to an effective Hamiltonian
\begin{equation}
  H = \frac{1}{2}\sum_n \Tr{\Pi_a^2(n)} + S_B[\cU, \cUbar] + \Phidag \left(\cDdag \cD\right)^{-1 / 4} \Phi.
\end{equation}
The core of the algorithm is molecular dynamics (MD) evolution of $\Pi_a$ and $\cU_a$ along a trajectory of length $\tau$ in a fictitious MD time.
Since the pseudofermion field already has the proper distribution, $\Phi$ is kept fixed along this trajectory.
The equation of motion for the links is simplified by the fact that they live in \glN rather than SU($N$).
The MD evolution involves inexact integration of Hamilton's equations, which does not conserve the effective Hamiltonian, $\De H \ne 0$.
The final step of the RHMC algorithm is a Metropolis--Rosenbluth--Teller test: the new gauge field values produced by the MD evolution are accepted with probability $P = \mbox{min}(1, e^{-\De H})$, otherwise the original field values are restored to produce a new configuration identical to the starting configuration.
So long as the MD integration scheme is area preserving (symplectic) and reversible (symmetric), the acceptance test makes the algorithm exact by stochastically correcting for the integration errors.

The above summary of RHMC is almost identical to that of the more familiar HMC algorithm~\cite{Duane:1987de}.
The main difference between the two appears in the MD evolution, each step of which requires solving
\begin{align}
  \label{eq:forces}
  \deriv{\cU_a}{t} & = \deriv{H}{\Pi_a} = \Pi_a &
  \deriv{\Pi_a}{t} & = -\deriv{H}{\cU_a} = -\deriv{S_B}{\cU_a} - \deriv{}{\cU_a} \Phidag \left(\cDdag \cD\right)^{-1 / 4} \Phi.
\end{align}
In order to take the Lie derivative of the pseudofermion term in the effective Hamiltonian, we approximate the rational power of the matrix inverse by a series of $P$ partial fractions,
\begin{equation}
  \label{eq:pfrac}
  \left(\cDdag \cD\right)^{-1 / 4} = \al_0 + \sum_{i = 1}^P \al_i \left(\cDdag \cD + \be_i\right)^{-1}
\end{equation}
(and similarly for the rational power $\left(\cDdag \cD\right)^{1 / 8}$ needed to initialize $\Phi$).
The fermionic contribution to the force $\deriv{H}{\cU_a}$ then becomes
\begin{equation}
  \deriv{}{\cU_a} \Phidag \left(\cDdag \cD\right)^{-1 / 4} \Phi = \sum_{i = 1}^P \al_i \left[(\cDdag \cD + \be_i)^{-1}\Phi\right]^{\dag}\deriv{\left(\cDdag \cD\right)}{\cU_a}\left[(\cDdag \cD + \be_i)^{-1}\Phi\right],
\end{equation}
and all the $(\cDdag \cD + \be_i)^{-1}\Phi$ can be efficiently determined by a multi-shift conjugate gradient (CG) inverter~\cite{Jegerlehner:1996pm}.
These inversions dominate the cost of lattice computations.

In Sections~\ref{sec:run} and \ref{sec:code} we will provide some further details about the rational approximations and MD integration schemes currently provided by \SLAT.
For now we remark that optimal shifts $\be_i$ and amplitudes $\al_i$ are computed offline by minimizing the relative approximation error within a given spectral range $[\la_{low}, \la_{high}]$, with $\la_{low} > 0$~\cite{Clark:Remez}.
Keeping the errors negligible across larger spectral ranges requires adding more terms to the partial fraction approximation, increasing computational costs.
At a minimum, we must demand that the smallest and largest eigenvalues of $\cDdag \cD$ fall within the spectral range we use, $\la_{low} < \la_{min} \ll \la_{max} < \la_{high}$.
The extremal eigenvalues of the squared fermion operator are therefore, like the Pfaffian, important quantities to monitor during RHMC evolution.
Fortunately such eigenvalue computations are much cheaper than Pfaffian measurements, and in \SLAT\ we carry them out using the PReconditioned Iterative Multi-Method Eigensolver (PRIMME) library~\cite{Stathopoulos:2010}.

By repeating the steps of the RHMC algorithm to accumulate some number (typically thousands) of MD time units, we generate an ensemble of gauge configurations.
We have already discussed some important observables to measure using these configurations, namely the Pfaffian and low-lying eigenvalues of the fermion operator.
A simpler set of common lattice observables consists of products of gauge links around closed loops on the lattice.
Since the links are elements of $\glN$, they are not unitarized and the simplest possible observable $\frac{1}{N}\Tr{\cU_a(x) \cUbar_a(x)}$ is non-trivial.
The plaquette $\Tr{\cU_b(x) \cU_a(x + \muhat_b) \cUbar_b(x + \muhat_a) \cUbar_a(x)}$ is also straightforward to compute even on the $A_4^*$ lattice with five basis vectors at each lattice site, and its determinant provides information about how far the links are from $\slN$.
The average link-trace, plaquette, Polyakov loop $\Tr{\prod_{N_t} \cU_t}$, and all contributions to the lattice action are measured and printed after every trajectory.

Further observables are typically computed only on stored configurations.
These include larger Wilson loops oriented along the principal axes of the lattice, which can be used to monitor the restoration of the $\cQ_a$ and $\cQ_{ab}$ supersymmetries, as discussed in \refcite{Catterall:2014vka}.
To monitor the \cQ supersymmetry that is broken only by non-zero $\mu$ and \ka in $S_{stab}$ (\eq{eq:Sstab}), we measure violations of the Ward identity
\begin{equation}
  \cQ \cO = \Tr{\sum_b \left(\cU_b\cUbar_b - \cUbar_b\cU_b\right) \sum_a \cU_a \cUbar_a} - \Tr{\eta \sum_a \psi_a \cUbar_a} = 0,
\end{equation}
which requires computing the fermion bilinear $\eta \psi_a$.
In \SLAT\ we compute $\eta \psi_a$ by inverting the fermion operator \cD on Gaussian-distributed stochastic (``noisy'') sources, a standard technique used to compute the chiral condensate in lattice QCD.\footnote{The structure of the fermion operator allows us to obtain two estimates of $\eta \psi_a$ for each stochastic source, by simultaneously computing both the $\eta$ components of $\cD^{-1}\psi_a$ and the $\psi_a$ components of $\cD^{-1}\eta$.}
The static potential is also extracted from Wilson loops, but in this case it is advantageous to measure loops for all possible spatial separations $\vec r$, not just the ``on-axis'' loops that lie along the principal axes of the lattice.
To do so, we gauge fix to Coulomb gauge and compute
\begin{equation}
  W(\vec r, t) = \Tr{P(\vec x, t, t_0) P^{\dag}(\vec x + \vec r, t, t_0)},
\end{equation}
where $P(\vec x, t, t_0)$ is a product of temporal links $\cU_t$ at spatial location $\vec x$, which extends from timeslice $t_0$ to timeslice $t_0 + t$.

\section{\label{sec:run}Running the code: High-level overview} 
In this section we summarize how to obtain \SLAT, compile the provided programs, and run the resulting executables on a generic GNU/Linux system.
In the next section we will describe the code itself in greater detail, providing information necessary to extend it with additional measurements or other features.

\SLAT\ is developed in a publicly accessible version control repository~\cite{GitHub}, so that one way to download the most recent version of the code is to run
\vspace{-6 pt}
\begin{verbatim}$ git clone https://github.com/daschaich/susy.git\end{verbatim}
\vspace{-6 pt}
Alternately, a tarball containing the version of the code discussed in this article may be found at \url{http://github.com/daschaich/susy/archive/arXiv.tar.gz}. 
After obtaining the code, move to the \verb!4dSYM/susy! application directory.
In addition to a \verb!README! and other files to be discussed in the next section, this directory contains two sample Makefiles: \verb!Make_scalar! uses \verb!gcc! to compile serial executables, while \verb!Make_mpi! uses \verb!mpicc! to compile parallel (MPI) executables.
These Makefiles may need minor modifications on a given system.
(For example, \verb!Make_mpi! is set up for the USQCD clusters at Fermilab.)
They are both wrappers that load architecture-independent information from the main \verb!Make_template!, where all the available executables are defined.

To compile RHMC evolution for serial (``scalar'') execution, simply run
\vspace{-6 pt}
\begin{verbatim}$ make -f Make_scalar susy_hmc\end{verbatim}
\vspace{-6 pt}
This will produce an executable named \verb!susy_hmc!, one of several executables that are actively maintained and well tested:
\vspace{-6 pt}
\begin{itemize}
  \setlength{\itemsep}{1 pt}
  \setlength{\parskip}{0 pt}
  \setlength{\parsep}{0 pt}
  \item \verb!susy_hmc! runs the RHMC algorithm for gauge configuration generation, along with inexpensive basic measurements (e.g., of the plaquette, link-trace and Polyakov loop) that can be used to monitor the RHMC evolution.
  \item \verb!susy_meas! carries out more expensive measurements on saved gauge configurations, as discussed at the end of the previous section.  These include both on-axis and gauge-fixed Wilson loop computations, with optional stout smearing.  Measurements of the $\eta \psi_a$ fermion bilinear from inversions using a given number of stochastic sources are also optional, and dominate the computational cost if performed.
  \item \verb!susy_eig! is an interface to PRIMME~\cite{Stathopoulos:2010} that computes the smallest and largest eigenvalues of $\cDdag \cD$ for a given gauge configuration, along with the corresponding eigenvectors.
  \item \verb!susy_phase! computes the complex Pfaffian of the fermion operator $\cD[\cU, \cUbar]$ for a given gauge configuration, using the algorithm described in \secref{sec:Pfaffian}.
\end{itemize}
An additional \verb!susy_hmc_meas! executable carries out the same measurements as \verb!susy_meas! in the course of RHMC evolution.

In the \verb!testsuite/scalar! directory we provide a script named \verb!run_tests! that will compile each of these executables, run it in serial using a fixed set of input options, and check the resulting output against reference files distributed with the code.
The \verb!run_tests! script takes two optional command-line arguments.
Running
\vspace{-6 pt}
\begin{verbatim}$ ./run_tests <N> <target>\end{verbatim}
\vspace{-6 pt}
will compile, run and check the \verb!susy_$target! executable for gauge group U(\verb!$N!), where \verb!$N! must be either 2, 3 or 4.
Alternately, if \verb!run_tests! is executed with no arguments it will check every executable listed above, and then test that the Pfaffian computation can be split into multiple checkpointed pieces.
Running every test in this manner will take several hours, with the bulk of the time spent computing the Pfaffian.
Similar tests of the parallel executables can be carried out by customizing the \verb!run_tests! script in the \verb!testsuite/mpi! directory to use the appropriate Makefile and \verb!mpirun! for the MPI system to be tested.
So long as the parallel tests are run using two cores, the reference output should be reproduced.

All regular \SLAT\ output is written to \verb!stdout!, and is typically redirected to a single plain-text file.
Each measurement appears on a line that starts with a corresponding label, which can easily be selected by offline analyses using awk, python, perl, etc.
The in-line documentation specifies the labels currently used, along with other details of the output.

In the next two subsections we will summarize various options that users may specify when using \SLAT.
Some of these options have to be fixed when compiling, while others are read in by the executables when they are run.
We will then conclude this section by discussing the scaling and parallel performance of \SLAT.

\subsection{\label{sec:compile}Compile-time options} 
\begin{figure}[htbp]
  \centering
  \includegraphics[height=\figheight]{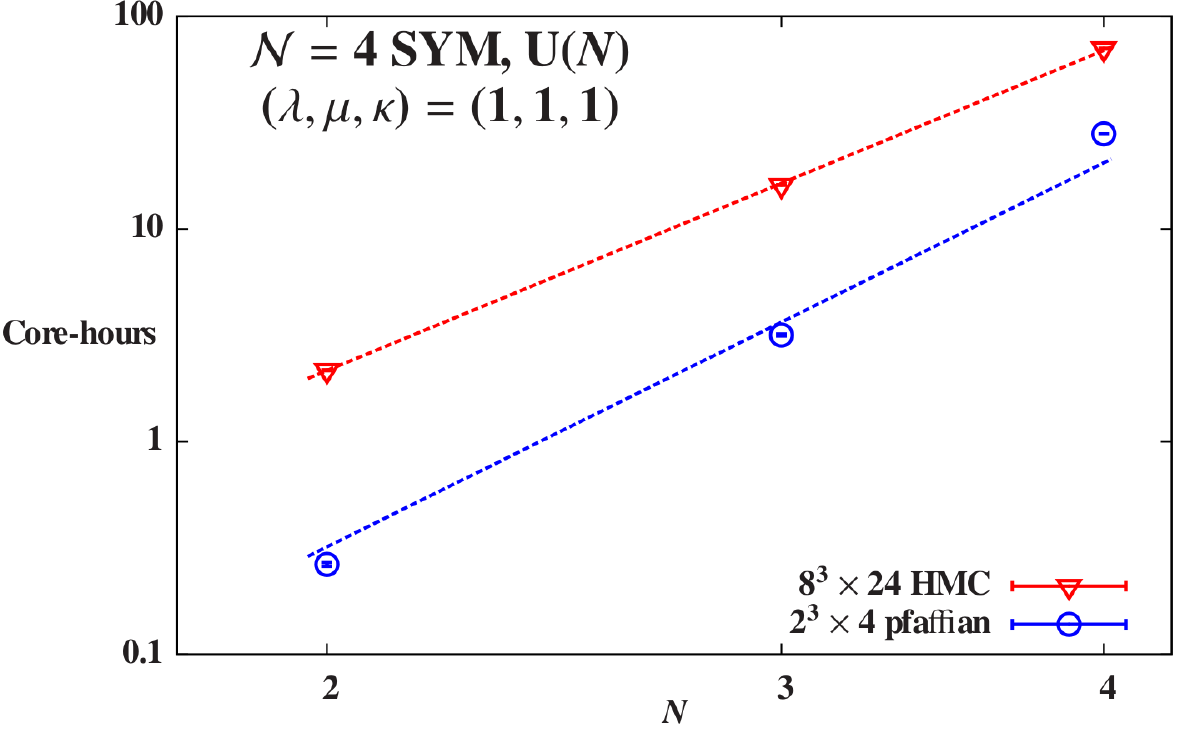}
  \caption{\label{fig:costs_N} Total computational costs for $8^3\X24$ RHMC generation of a single molecular dynamics time unit, and $2^3\X4$ Pfaffian measurements, for gauge groups U(2), U(3) and U(4), on log--log axes.  The red dashed line illustrates the cost scaling $\propto N^5$ observed for RHMC configuration generation, while the blue dashed line shows the expected $N^6$ scaling for Pfaffian measurements.}
\end{figure}

The executables listed above are produced by compiling with appropriate combinations of macro definitions: \verb!-DHMC_ALGORITHM! for RHMC evolution; \verb!-DWLOOP! for gauge-fixed Wilson loop computations; \verb!-DSTOUT! for stout smearing; \verb!-DBILIN! for fermion bilinear measurements using stochastic sources; \verb!-DEIG! for eigenvalue measurements using PRIMME; and \verb!-DPHASE! to compute the complex Pfaffian.
In addition, several other features of the code must be fixed at the time of compilation.
For example, the gauge group U($N$) is chosen by uncommenting one (and only one) definition of the number of colors $\NCOL = N$ and corresponding dimension of the adjoint fermion representation $\DIMF = \NCOL^2$ at the top of the header file \verb!include/su3.h!.
The default gauge group is U(2).
Gauge groups U(3) and U(4) are also well tested and available through this header file.
Larger values of \NCOL\ require further extensions of the code, as we will discuss in \secref{sec:NCOL}.
In practice, computational costs increase rapidly as the number of colors grows.
In \fig{fig:costs_N}, we observe the costs of RHMC configuration generation increasing $\propto N^5$, while the Pfaffian calculation scales $\propto \DIMF^3 = \NCOL^6$ as expected.
This behavior makes it unlikely that computations with $N \geq 5$ will be practical in the immediate future.
Even so, because deviations from the large-$N$ limit of $\cN = 4$ SYM go as $1 / N^2$, investigations with $N \leq 4$ should access the large-$N$ regime up to few-percent effects that may be comparable to statistical uncertainties.

Other important compile-time options may be found in the header file \verb!susy/defines.h!.
The simplest of these is \verb!PBC!, which controls the temporal boundary conditions (BCs) of the fermions.
This flag can be set to 1 for periodic BCs, or $-1$ for antiperiodic (thermal) BCs.
The default antiperiodic BCs are recommended, to lift a fermionic zero mode that may otherwise destabilize the computations.
The coefficient $c_2$ in $S_{exact}$ (\eq{eq:Sexact}) may be changed by redefining the flag \verb!C2! in this file.
We recommend leaving this coefficient set to its classical value $c_2 = 1$.

Finally, the order $P$ of the rational approximation in \eq{eq:pfrac} is set by the flag \verb!DEGREE! in \verb!defines.h!.
Five options are supported, each of which provides rational approximations that reproduce $\left(\cDdag \cD\right)^{-1 / 4}$ and $\left(\cDdag \cD\right)^{1 / 8}$ up to fractional errors of at most $2\times 10^{-5}$ across a given spectral range:
\begin{itemize}
  \setlength{\itemsep}{1 pt}
  \setlength{\parskip}{0 pt}
  \setlength{\parsep}{0 pt}
  \item \verb!DEGREE! 5 is set up for the spectral range $[0.1, 50]$
  \item \verb!DEGREE! 6 is set up for the spectral range $[0.02, 50]$
  \item \verb!DEGREE! 8 is set up for the spectral range $[10^{-3}, 50]$
  \item \verb!DEGREE! 9 is set up for the spectral range $[10^{-4}, 45]$
  \item \verb!DEGREE! 15 is set up for the spectral range $[10^{-7}, 1000]$
\end{itemize}
A smaller \verb!DEGREE! allows the CG inverter to converge more quickly, but if the eigenvalues of $\cDdag \cD$ exceed the spectral range of the rational approximation, significant numerical errors may result.
A \verb!DEGREE! of 8 or less suffices for our typical computations.

\subsection{Run-time options} 
When run, each executable must be given a set of input parameters that specify the lattice volume, the gauge configuration to load, and other pertinent information.
All possible input parameters for every executable discussed above are summarized in \fig{fig:sample_input}.
Because this figure combines input options for multiple executables, it cannot be used as an input file itself.
Sample input files for each executable are provided in the \verb!testsuite! directory.
Although comments and whitespace are ignored, the options must come in the order expected by the executable, and each variable must be immediately preceded by its name as shown in \fig{fig:sample_input}.

\begin{figure}[htp]
  \centering
  \caption{\label{fig:sample_input} All possible input parameters that may be read in by a \SLAT\ executable, with brief in-line explanations.}
  \fbox{
\begin{lstlisting}[language=bash, firstnumber=0]
prompt 0  # If non-zero, prompt user to manually input each parameter
nx 6      # Lattice volume is nx*ny*nz*nt
ny 6
nz 6
nt 6
iseed 41  # Random number generator seed

warms 0               # Number of trajectories without expensive measurements
trajecs 10            # Number of trajectories with expensive measurements
traj_length 1         # Trajectory length (tau)
nstep 10              # Fermion steps per trajectory; step_size = traj_length / nstep
nstep_gauge 10        # Gauge steps per fermion step
traj_between_meas 10  # How many trajectories to skip between expensive measurements

lambda   1.0  # 't Hooft coupling
kappa_u1 0.5  # Plaquette determinant coupling
bmass    0.5  # Bosonic mass parameter (mu)
fmass    0.0  # Mass shift in fermion operator

# The next two lines must only be included when compiling with -DSTOUT
Nstout 1  # Number of stout smearing steps
rho 0.1   # Stout smearing parameter

max_cg_iterations 5000  # Maximum number of CG iterations in each inversion
error_per_site 1e-5     # "Linear" CG stopping condition (squared to compare with |r|^2)

# The next line must only be included when compiling with -DBILIN
nsrc 3    # Number of stochastic sources for fermion bilinear calculations

# The next three lines must only be included when compiling with -DEIG
Nvec 100      # Number of eigenvalues and eigenvectors to calculate
eig_tol 1e-8  # Eigensolver tolerance (convergence criterion)
maxIter 5000  # Maximum number of eigensolver iterations

# The next two lines must only be included when compiling with -DPHASE
ckpt_load -1  # If positive, load checkpointed pfaffian computation
              # from config.Q$ckpt_load and config.diag$ckpt_load
ckpt_save -1  # If positive, checkpoint pfaffian computation
              # to config.Q$ckpt_save and config.diag$ckpt_save

reload_serial config    # How to set up lattice at start:
                        # only fresh, random, continue and reload_serial supported
no_gauge_fix            # Only coulomb_gauge_fix and no_gauge_fix supported
save_serial new-config  # What to do with lattice at end: only forget and save_serial supported
\end{lstlisting}}
\end{figure}

While the in-line explanations in \fig{fig:sample_input} should suffice to explain most options, some further comments are called for.
First, the lines starting with \verb!reload_serial!, \verb!warms!, \verb!trajecs! and \verb!save_serial! determine the overall structure of an RHMC computation.
In this case, the configuration \verb!config! would be loaded from disk, the RHMC algorithm would be run for ten trajectories (with trajectory length of $\tau = 1$ MD time unit between accept/reject tests set by \verb!traj_length!), and the resulting configuration would be saved to disk as \verb!new-config!.
A trajectory length of $\tau = 1$ is a reasonable default.
Although longer trajectories decrease autocorrelations between subsequent configurations in the Markov chain, they also require smaller steps in the MD evolution, to keep $|\De H| \lesssim 1$ and maintain reasonable acceptance.
The details of optional checkpointing in the Pfaffian computation will be discussed in \secref{sec:Pfaffian}.
Finally, we emphasize that the \verb!fmass! parameter in the input is {\em not} the $\cQ$-invariant coefficient $m$ that appears in \eq{eq:fmass}.
Instead this parameter simply shifts the squared fermion operator
\begin{equation}
  \cDdag \cD \lra \cDdag \cD + \verb!fmass!^2.
\end{equation}
This shift is not needed when the fermionic fields are subject to antiperiodic temporal BCs, and we recommend using compile-time option \verb!PBC = -1! and \verb!fmass = 0!.

\subsection{Performance and scaling} 
The performance advantages from parallelization are dramatic, and greatly extend the applicability of \SLAT\ compared to the serial code presented in \refcite{Catterall:2011cea}.
For typical small-volume calculations (e.g., $4^3\X12$ gauge configuration generation with $N = 2$), \SLAT\ is roughly two orders of magnitude faster than the serial code when each is run in its standard production environment (including single-GPU acceleration of the serial code's multi-shift CG inverter~\cite{Galvez:2011cd}, which significantly improves its performance). 
More importantly, \SLAT\ can be scaled up to larger lattice volumes and larger gauge groups than the serial code can handle -- the restriction to a single GPU limits the serial code to lattice volumes of at most $8^3\X16$ even for the smallest $N = 2$.

\begin{figure}[bp]
  \centering
  \includegraphics[height=\figheight]{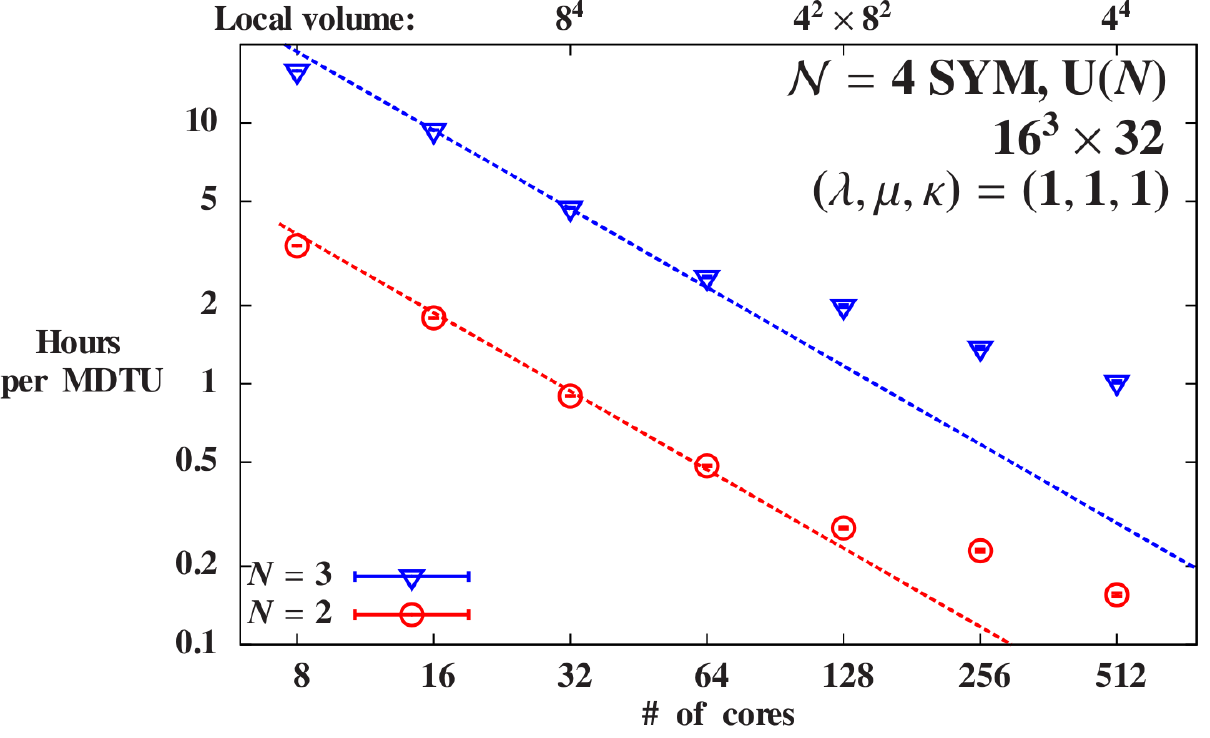}
  \caption{\label{fig:RHMC_scaling} Strong scaling of the time to solution for U(2) and U(3) \sixt RHMC generation of a single molecular dynamics time unit on the USQCD \pizero\ cluster at Fermilab, on log--log axes.  The dashed lines illustrate optimal scaling (a power of $-1$), which the computations follow fairly well through 128 cores (a local volume of $4^2\X8^2$ lattice sites per core), with deteriorating performance on 256 and 512 cores (local volumes $4^3\X8$ and $4^4$).}
\end{figure}

Efficiently studying larger lattice volumes and larger $N$ requires that \SLAT\ performs well when running across many cores.
We observe good parallel scaling from \SLAT\ (thanks in large part to the work done over many years to optimize the MILC code), although further performance improvements may be feasible.
First consider the strong scaling shown in \fig{fig:RHMC_scaling}, which considers the time to solution for a fixed computation spread across various numbers of cores.
The computation under consideration is to generate a single MD time unit with the RHMC algorithm, for a $\sixt$-site lattice volume and either gauge group U(2) or U(3).
The time to solution steadily decreases as we use up to 512 cores (32 nodes) of the USQCD \pizero\ cluster at Fermilab.

In particular, up to around 128 cores (eight 16-core \pizero\ nodes) the scaling is roughly optimal, following a straight line with slope $-1$ on the logarithmic axes of \fig{fig:RHMC_scaling}.
The local volume on 128 cores is $4^2\X8^2$ lattice sites per core, which decreases to $4^3\times 8$ and $4^4$ sites per core on 256 and 512 cores, respectively.
The performance deteriorates on more than 128 cores, as communication between all these cores begins to take longer than the computations that each core has to carry out.
Although lattice $\cN = 4$ SYM involves more computation per site than lattice QCD with staggered or Wilson quarks, it also requires much more intercore communication.
In particular, the structure of the $S_{closed}$ term in \eq{eq:Sclosed} introduces dozens of intercore data transfers (``gathers'') into each application of the fermion operator (the matrix--vector operation in the CG algorithm).

\begin{figure}[bp]
  \centering
  \includegraphics[height=\figheight]{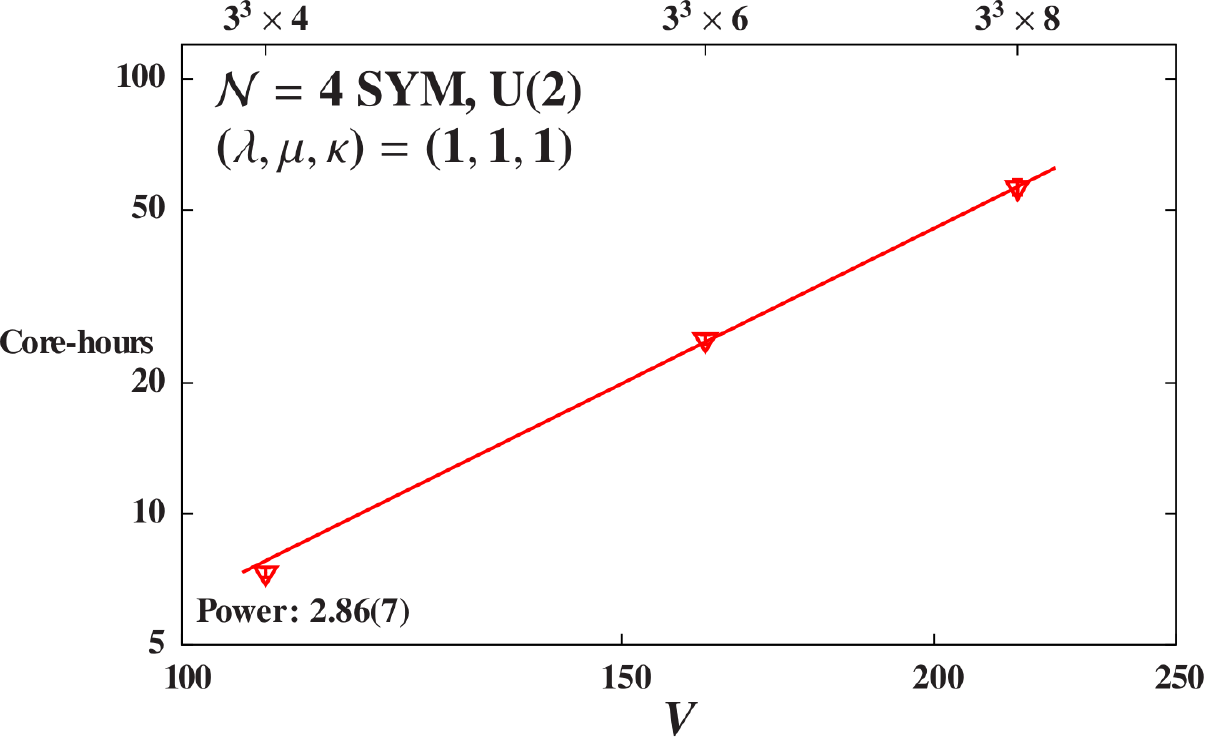}
  \caption{\label{fig:pfaffian_scaling} Weak scaling of the total computational cost for U(2) Pfaffian measurements with fixed local volume $3^3\X2$, on log--log axes with a power-law fit.  Since the Pfaffian computation requires $\cO(V^3)$ operations, the power of approximately 3 indicates optimal scaling of the code.}
\end{figure}

We can also consider weak scaling, in which the problem size increases in tandem with the number of cores being used.
In \fig{fig:pfaffian_scaling} we plot the total computational cost in core-hours for Pfaffian measurements with gauge group U(2) and lattice volumes $3^3\X4$, $3^3\X6$ and $3^3\X8$ on 2, 3 and 4 cores, respectively.
That is, the local volume of $3^3\X2$ sites per core is fixed, and all computations use part of a single 8-core node in the HEP-TH cluster at the University of Colorado.
The straight line on the logarithmic axes of \fig{fig:pfaffian_scaling} is a fit to power-law scaling, with a power of 2.86(7).
The Pfaffian computation requires $\cO(V^3)$ operations, so this power of approximately 3 indicates optimal weak scaling of the code.

\section{\label{sec:code}Modifying the code: Underlying details} 
In this section we provide more information about elements of the code itself, highlighting notable aspects of the documentation for the benefit of those who may be interested in contributing to future development.
\SLAT\ evolved from a version of the MILC QCD code that was extended by DeGrand, Shamir and Svetitsky (DSS) to handle SU($N$) gauge groups with arbitrary $N$ and fermions transforming in either the fundamental, adjoint, two-index symmetric or two-index antisymmetric representations.
Compared to the DSS code, further changes were required to add the fifth link required by the $A_4^*$ lattice structure, to implement \KD fermions and the $\cN = 4$ SYM lattice action, and to convert the gauge links from SU($N$) to $\glN$.
Even so, the basic structure of the code continues to resemble MILC, and much of the information in the MILC manual remains relevant~\cite{MILC:manual}.
After describing the directory tree, in the subsequent subsections we will address issues that make \SLAT\ different from a typical parallel QCD code.

The five subdirectories within the \verb!4dSYM! directory are set up analogously to the MILC code:
\vspace{-6 pt}
\begin{itemize}
  \setlength{\itemsep}{1 pt}
  \setlength{\parskip}{0 pt}
  \setlength{\parsep}{0 pt}
  \item \verb!libraries! contains simple single-processor routines, for example $N\X N$ matrix multiplication.
  \item \verb!include! holds header files for procedures common to typical applications, as well as important compile-time definitions such as the number of colors $N$ and basic $N$-dependent data structures.
  \item \verb!generic! implements procedures common to typical applications, such as data layout, internode communications, and input/output (I/O).
  \item \verb!susy! is currently the only application directory, containing code for four-dimensional $\cN = 4$ SYM.
  \item \verb!testsuite! provides reference input and output files for all actively maintained executables, along with scripts to test any or all of these executables, in serial or in parallel, for U($N$) gauge theory with $N = 2$, 3 or 4 colors.
\end{itemize}
The \verb!2dSYM! directory with the same structure is set up for two-dimensional $\cN = (2, 2)$ SYM theory, which we briefly discuss in the appendix.

\subsection{\label{sec:NCOL}Gauge group and fermion representation} 
To generalize the MILC code to handle arbitrary SU($N$), the DSS code replaced the hardwired ``\verb!3!'' for the number of colors in SU(3) with a macro definition \NCOL.
This involved modifications to most single-processor library routines (e.g., $\NCOL \X \NCOL$ matrix multiplication).
The bosonic fields in \SLAT\ are $\NCOL \X \NCOL$ complex matrices, just as the gauge fields in non-supersymmetric SU(\NCOL) lattice gauge theory would be.
However, because the lattice $\cN = 4$ SYM link variables are elements of \glN rather than SU($N$), they are not unitarized in the course of the simulation.

The DSS code allows for fermions transforming in representations other than the fundamental.
The dimensionality of the fermion representation is set by the macro definition \DIMF.
While fermions in the adjoint representation can be naturally encoded in a set of $\NCOL \X \NCOL$ complex matrices, the DSS code remains more general by making the fermionic fields \DIMF-component vectors in color space.
We retain this setup in \SLAT, with $\DIMF = \NCOL^2$ fixed for the U(\NCOL) adjoint representation.
This approach forces us to keep two sets of bosonic variables in the code.
One set, called \verb!su3_matrix_f!, denotes $\NCOL \X \NCOL$ complex matrices and is used in the bosonic part of the code.
The other set, called \verb!su3_matrix!, denotes $\DIMF \X \DIMF$ matrices and appears in the routines involving (pseudo)fermions, such as the calculation of the pseudofermion action through \eq{eq:pf}.
An \verb!su3_matrix! $\cV_a$ is composed from an \verb!su3_matrix_f! $\cU_a$ by
\begin{equation}
  \cV_a^{AB} = \Tr{\la_{ij}^A \left(\cU_a\right)_{jk} \la^B_{ki}},
\end{equation}
where the $\la^A$ matrices are generators of U($N$) defined in the file \verb!susy/setup_lambda.c!.
Whenever the gauge configuration $\cU_a$ is updated, the ``adjoint links'' $\cV_a$ must be regenerated by the \verb!fermion_rep()! routine.
Of course, one must keep a complete set of library routines for matrix operations involving either \verb!su3_matrix! or \verb!su3_matrix_f! variables.

As mentioned in \secref{sec:compile}, certain structures and routines are currently implemented only for $\NCOL \leq 4$.
These include the \verb!anti_hermitmat! structure in \verb!include/su3.h!, which explicitly lists the $\NCOL(\NCOL + 1) / 2$ non-trivial components of $\NCOL \X \NCOL$ anti-hermitian complex matrices.
This structure is used for stout smearing, by routines that set up random gauge configurations, and by four library routines.
In addition, monitoring the restoration of the $\cQ_a$ and $\cQ_{ab}$ supersymmetries requires inverting $\NCOL \X \NCOL$ gauge link matrices, which is implemented using the cofactor matrix for $\NCOL \leq 4$.
While it is straightforward to extend these structures and routines to $\NCOL \geq 5$, the rapidly growing computational costs make such work unlikely to prove useful at present.

\subsection{Data layout and communications} 
The MILC code lays out its dynamical variables in two ways.
``Site-wise'' variables are packaged into a \verb!site! structure defined globally in the header file \verb!susy/lattice.h!.
The lattice itself is an array of such \verb!site!s, each of which contains, inter alia, five gauge links $\cU_a$ and five adjoint links $\cV_a$.
This construction dates from the earliest version of the MILC code, when this layout was optimal for the compilers and computing architectures then in use.
More recently the trend has been to define each lattice variable as a separate ``field-wise'' array assigned a contiguous block of memory on each node; see Section~5.3 of \refcite{MILC:manual} for further discussion.
These field-wise variables are also globally defined in \verb!susy/lattice.h!, and allocated in \verb!susy/setup.c! after the lattice volume is set at run time (\fig{fig:sample_input}).

Because both site-wise and field-wise variables coexist in \SLAT, we must maintain two sets of internode communication (``gather'') routines, with each set handling either site-wise or field-wise data layout.
In addition, the macro \verb!FORALLSITES! is provided to simplify looping over the lattice volume.
This macro keeps track of both the linear index of field-wise variables along with a pointer to the corresponding \verb!site!, making it easy to use the two types of variables in combination.

A new structure in \SLAT\ that does not appear in the MILC code is called \verb!Twist_Fermion!, and bundles together all the fermion variables $\Psi = (\eta, \psi_a, \chi_{ab})$, for convenience.

The MILC code was written to deal with four-dimensional gauge fields discretized on hypercubic lattices.
Even using the convenient representation of the $A_4^*$ lattice discussed in \secref{sec:lattice}, in which four of the five gauge links at each site are the usual hypercubic objects, accommodating the fifth link $\muhat_4 = (-1, -1, -1, -1)$ required rewriting various low-level routines.
For example, utilities for reading and writing binary gauge configuration files had to be extended to include the fifth link.

In addition, in the MILC code each \verb!site! in the lattice carries a flag labeling its parity in the sense of an even/odd checkerboard: the parity is
\begin{equation}
  p(n) = \mbox{mod}\left(\sum_{i = 0}^3 n_i, 2\right).
\end{equation}
The first four links connect ``even'' ($p = 0$) sites to ``odd'' ($p = 1$) sites, and vice versa.
This checkerboarding is widely used in ordinary QCD codes.
For example, staggered fermions are often defined to live on only one parity, both at the level of HMC evolution and in measurements.
In \SLAT\ it is hard to use checkerboarding, since the fifth link connects sites of the same parity.

The only context in which we make use of the parity flag is parallel gauge fixing for static potential analyses.
A typical gauge choice is one which maximizes some set of links.
The only option currently implemented in \SLAT\ is Coulomb gauge, which maximizes $\sum_{i = 0}^2 \Tr{\cU_i(n)}$ at every site.
The formula for the gauge transformation, $\cU_a(n) \to G(n) \cU_a(n) \Gdag(n + \muhat_a)$, motivates considering an iterative sequence of gauge transformations $G$ that are alternately taken to be the identity on even or on odd sites.
Then for the first four links the maximization becomes linear in the $G$s, reducing to multiplication on the left for links at sites of the parity for which $G$ is nontrivial, and on the right for links at sites of the other parity.
The fifth link always has to be multiplied on both sides, and cannot be included in the maximization without substantially rewriting the gauge fixing code.
This is not an issue for Coulomb gauge, which includes only the first three links in the gauge condition.

Next, we must implement antiperiodic temporal BCs for the fermions differently than in standard QCD codes where matter fields are defined only on the sites.
In the QCD context, such BCs can be included simply by negating the temporal gauge links on the last timeslice of the lattice.
Since the $\psi_a$ and $\chi_{ab}$ fermionic variables in \SLAT\ connect different sites, this trick doesn't work.
Instead, we include in each \verb!site! the necessary factors of $\texttt{PBC} = \pm1$ (discussed in \secref{sec:compile}) to be applied to the fermion fields whenever they cross the temporal boundary.\footnote{We are free to take any of the five links to define the temporal direction, so we use the same $\muhat_3 = (0, 0, 0, 1)$ as MILC.}
This is the most straightforward implementation of antiperiodic temporal BCs, but is unlikely to be the most efficient.
These BC factors are set up in \verb!susy/setup_offset.c! after the lattice has been laid out.

The primary purpose of this \verb!setup_offset.c! file is to speed up the internode communications mentioned above.
Both site-wise and field-wise gathers come in two varieties.
The usual gathers can move data only along specific paths that are tracked by a global static \verb!gather_array! set up by routines in \verb!generic/com_mpi.c!.
For example, these include the paths connecting nearest-neighbor sites in the four spacetime directions.
In contrast, ``general'' gathers handle an arbitrary lattice displacement four-vector, but take much longer to run.
Every general gather must determine which nodes will exchange data, while this information is precomputed for the usual gathers.
In addition, only one general gather can run at a time.
Many of the usual gathers may be carried out simultaneously, which we use to overlap communications with computations.

In order to use the usual gathers even for the several dozen two- and three-link paths required by the $\cQ$-closed part of the action (\eq{eq:Sclosed}), in \verb!susy/setup_offset.c! we add all of these paths to the \verb!gather_array!.
We also add all five $A_4^*$ link paths to the \verb!gather_array!, in addition to the four spacetime directions set up by default in \verb!generic/com_mpi.c!.
We keep track of all these new gathers in the arrays \verb!gq_offset! and \verb!goffset!, respectively.
Although the first four $A_4^*$ link paths duplicate the four spacetime directions, the latter are set up to allow even/odd checkerboarding as in MILC, while the former are not.
In the rare event that checkerboarding is needed, the standard MILC-like gather paths must be used.
Otherwise we recommend gathering via the \verb!goffset! array.

Avoiding general gathers in the $\cQ$-closed term significantly improves the performance of \SLAT, by roughly a factor of two.
Overlapping communications with computations in this term is also beneficial when running on many nodes.
For U(2) $12^4$ gauge generation on 64 cores (2 nodes) of the USQCD \bc\ cluster at Fermilab, with local volume $3^2\X 6^2$, such overlapping reduced costs by an additional $\sim$30\%.

\subsection{Rational hybrid Monte Carlo} 
As summarized in \secref{sec:RHMC}, the molecular dynamics (MD) evolution at the heart of the RHMC algorithm is responsible for the bulk of computing in typical lattice calculations.
Lattice QCD researchers have developed many clever techniques to make the MD evolution more efficient, several of which we import to \SLAT\ from the DSS code.
Our default MD algorithm is a second-order Omelyan integrator~\cite{Omelyan:2002PRE, Omelyan:2003CPC}, with separate time scales~\cite{Urbach:2005ji} for the gauge and fermion force terms in \eq{eq:forces}.
The average gauge and fermion forces along each trajectory are printed in the output to help tune run parameters.

While developing \SLAT, it was useful first to implement a simple leapfrog integrator, against which our current algorithm could be tested.
In case similar testing is necessary in the future, we have retained this leapfrog integrator in the source code distribution.
It can be used by replacing \verb!update_o.o! with \verb!update_leapfrog.o! in \verb!susy/Make_template!.
The leapfrog integrator has only a single time scale, and for simplicity this is set by \verb!nstep! in \fig{fig:sample_input}, ignoring \verb!nstep_gauge!.

The five different rational approximations discussed in \secref{sec:compile} are set up in \verb!susy/setup_rhmc.c!, simply by copying in output from the Remez algorithm implemented in \refcite{Clark:Remez}.
For a given spectral range, we find the smallest degree $P$ in \eq{eq:pfrac} that produces a maximum fractional approximation error less than $2\X 10^{-5}$, a number set in previous work~\cite{Catterall:2011cea}.
While it is straightforward to add more approximations, those already available should provide good efficiency for foreseeable computations.
The shifts $\be_i$ and amplitudes $\al_i$ chosen at compile time are used by a standard multi-shift CG inverter~\cite{Jegerlehner:1996pm}.
We make it possible to redefine $P$, the shifts and the amplitudes during the computation, so that the same inversion routine also provides the basic CG used to compute the fermion bilinear $\eta \psi_a$.

\subsection{\label{sec:Pfaffian}Parallel Pfaffian computation} 
Computing the Pfaffian of the fermion operator \cD is a notoriously hard problem, but is necessary to assess the potential sign problem in our system.
In this section we present the algorithm we use to compute the Pfaffian in parallel, and discuss the checkpointing that allows these measurements to be split up into a sequence of computations.

The Pfaffian is formally
\begin{equation}
  \label{eq:Pfaffian}
  \pf \cD \equiv \int[d\Psi] \exp\left[-\Psi_i \cD_{ij}\Psi_j\right] = \frac{1}{N! 2^N} \eps_{\al_0\be_0\cdots\al_{N - 1}\be_{N - 1}} \cD_{\al_0\be_0}\cdots \cD_{\al_{N - 1}\be_{N - 1}},
\end{equation}
where the indices run over the lattice volume, the 16 fermion fields $\Psi = \left(\eta, \psi_a, \chi_{ab}\right)$, and the $N^2$ generators of U($N$), so that each $\Psi_i$ is a Grassmann variable.
The Pfaffian is only well-defined if \cD is an antisymmetric matrix of even dimension, which is the case for our fermion operator.
Up to a sign, about which we care a great deal, the Pfaffian is the square root of the determinant, $\det \cD = \left[\pf \cD\right]^2$.

We can compute the Pfaffian by considering the analogue of LU factorization for antisymmetric matrices, $\cD = L T L^T$, where $L$ is lower-triangular and $T$ is a trivial ($\pf T = 1$) tridiagonal matrix composed of $2\X2$ blocks $T_{i, i \pm 1} = \pm 1$.
As a consequence,
\begin{equation}
  \pf \cD = \pf \left[L T L^T\right] = \left(\det L\right) \pf T = \det L = \prod_i L_{ii}.
\end{equation}
There exist several algorithms to compute $\pf \cD$ based on this factorization~\cite{Campos:1999du, Catterall:2003ae, Wimmer:2012ae, Rubow:2011dq}.
All scale as $\cO(N_{\Psi}^3)$, where $N_{\Psi} = 16 N^2 L^3 N_t$ is the (even) number of elements in the fermionic fields for U($N$) gauge theory on an $L^3\X N_t$ lattice.

The algorithms described in Refs.~\cite{Wimmer:2012ae, Rubow:2011dq} aim to minimize the number of floating-point operations required to determine the Pfaffian.
Our main concern is to parallelize the computation, and in this context it is easier to consider the algorithms in Refs.~\cite{Campos:1999du, Catterall:2003ae}.
These employ procedures analogous to Gram--Schmidt orthogonalization in $2\X2$ blocks, based on applying the fermion operator itself, for which we already possess an efficient parallel implementation.

Specifically, the algorithm in the appendix of \refcite{Catterall:2003ae} constructs the upper-triangular matrix $Q = L^{-1}$ for which
\begin{equation}
  \label{eq:factor}
  Q \cD Q^T = T \Lra \left(\det Q\right) \pf \cD = 1 \Lra \pf \cD = \left(\det Q\right)^{-1} = \left(\prod_i Q_{ii}\right)^{-1}.
\end{equation}
The construction begins by initializing $Q$ to the unit matrix, $q^{(i)} = e_i$, where $q^{(i)}$ denotes the $i$th column of $Q$ and $e_i$ is the $i$th basis vector.
We cycle over pairs of columns, using the even column $q^{(i)}$ to normalize the odd column $q^{(i + 1)} \to q^{(i + 1)} / \left(q_a^{(i + 1)} \cD_{ab} q_b^{(i)}\right)$.
We then propagate each pair of columns through all subsequent columns $j \geq i + 2$,
\begin{equation}
  q^{(j)} \to q^{(j)} - q^{(i)} \left(q_a^{(i + 1)} \cD_{ab} q_b^{(j)}\right) + q^{(i + 1)} \left(q_a^{(i)} \cD_{ab} q_b^{(j)}\right)
\end{equation}
in order to obtain $Q \cD Q^T = T$.
In components, this condition is
\begin{equation}
  \left(q_a^{(i)} \cD_{ab} q_b^{(j)}\right) = \left\{\begin{array}{cl}1 & \mbox{ if } i = j + 1 \mbox{ and } j \mbox { is even} \\
                                                                     -1 & \mbox{ if } i = j - 1 \mbox{ and } j \mbox{ is odd} \\
                                                                      0 & \mbox{ otherwise}\end{array}\right. ,
\end{equation}
which we have explicitly checked for the matrix $Q$ produced by this algorithm.
(This test has since been removed from \SLAT, since it is incompatible with the optional checkpointing discussed below.)
Note that (as in \eq{eq:Pfaffian}) there is no complex conjugation in any of the inner products $\left(q_a^{(i)} \cD_{ab} q_b^{(j)}\right)$.
We retain the indices $a$ and $b$ being summed to emphasize this.

Algorithm~1 restates the description in the previous paragraph more compactly.
\begin{algorithm}
  \caption{Construction of upper-triangular $Q$ with $\det Q = \prod_i Q_{ii} = \left(\pf \cD\right)^{-1}$}
  \begin{algorithmic}
    \STATE $Q \gets \diag\left\{1, 1, \cdots, 1\right\}$
    \FOR{$i = 0$ to $N - 2$ by 2}
      \STATE $q^{(i + 1)} \gets q^{(i + 1)} / \left(q_a^{(i + 1)} \cD_{ab} q_b^{(i)}\right)$
      \FOR{$j = i + 2$ to $N - 1$ by 1}
        \STATE $q^{(j)} \gets q^{(j)} - q^{(i)} \left(q_a^{(i + 1)} \cD_{ab} q_b^{(j)}\right) + q^{(i + 1)} \left(q_a^{(i)} \cD_{ab} q_b^{(j)}\right)$
      \ENDFOR
    \ENDFOR
  \end{algorithmic}
\end{algorithm}
Obviously this algorithm will break down if $\left(q_a^{(i + 1)} \cD_{ab} q_b^{(i)}\right) = 0$ for any even column $q^{(i)}$.
Although this issue was not encountered in studies of the two-dimensional Wess--Zumino model considered by \refcite{Catterall:2003ae}, it can occur for lattice $\cN = 4$ SYM in four dimensions.
In order to guarantee that none of these inner products vanish, we must choose an appropriate basis for the matrices \cD and $Q$, which is determined by the order we cycle over the fermionic field components.
In particular, different $\eta^i$, $\psi_a^i$ and $\chi_{ab}^i$ must be considered on each site, before looping over the $i = 0, \cdots, N^2 - 1$ generators of U($N$).
For the two-dimensional $\cN = (2, 2)$ SYM theory discussed in the appendix, this step suffices to make all matrix elements non-zero.
For $\cN = 4$ SYM we also have to gather three $\chi_{ab}^i$ components from different lattice sites, instead of cycling over all the fields on a given site.
These extra gathers are a fairly small addition to the cost of each application of the fermion operator $\cD$.

Up to this point it has gone without saying that the sparse $N_{\Psi}\X N_{\Psi}$ matrix \cD is never explicitly constructed or stored in memory.
Instead, its action on a fermion vector (the matrix--vector or ``matvec'' operation) is implemented by the \verb!fermion_op! routine in \verb!susy/utilities.c!.
This is standard practice in lattice gauge theory.
In contrast, we must allocate the upper-triangular matrix $Q$ in order to compute it through Algorithm~1, even though we only care about the resulting diagonal components $Q_{ii}$.
The two nested loops in Algorithm~1 require $\left(N_{\Psi} / 2\right)^2$ matvec operations to compute $\cD q^{(i)}$ and $\cD q^{(j)}$ for steadily changing $q^{(j)}$.
The cost of each matvec operation is proportional to $N_{\Psi}$, producing the expected $\cO(N_{\Psi}^3)$ scaling of the overall computation.

For example, on a small $4^3\X6$ lattice with gauge group U(2), $N_{\Psi} = 24,576$ and Algorithm~1 requires roughly 151~million matvec operations.
We minimize the time to solution in this case by running on 16 of the 32 cores on a single node in the USQCD \ds\ cluster at Fermilab, resulting in a local volume of $2^3\X3$ sites per core.
This setup computes 217 matvecs per second on average, leading to a total runtime of about 8 days.

To avoid running such long jobs, we allow the Pfaffian computation to be checkpointed, by saving the partially-computed $Q$ to disk and reloading it in a subsequent job.
Checkpointing is controlled by the \verb!ckpt_save! and \verb!ckpt_load! options in \fig{fig:sample_input}.
Since $Q$ requires a large amount of memory (about 10~GB in the $4^3\X6$ case discussed above), we reduce disk space usage by discarding each column of $Q$ after its diagonal element $Q_{ii}$ has been determined through Algorithm~1.
Both the known diagonal elements and remaining columns that still need to be considered (specified by \verb!ckpt_save! and \verb!ckpt_load!) are saved to disk in separate scratch files.
In addition, we save only the non-zero elements of each column, which significantly reduces disk usage during the early stages of the computation.
This special I/O is implemented in \verb!generic/io_phase.c!.

\section{\label{sec:end}Summary and discussion of potential future developments} 
In this paper we have presented new parallel software for lattice studies of four-dimensional $\cN = 4$ SYM theory with gauge group SU($N$).
This system has several unusual features compared to more familiar QCD-like lattice gauge theories, in particular the non-hypercubic $A_4^*$ lattice with five basis vectors symmetrically spanning four dimensions, the \KD fermions spread out across both sites and links, and the complexified gauge links living in $\glN$, requiring approximate projection to the target SU($N$) gauge theory.
We reviewed the main elements of the system in \secref{sec:physics}, also discussing how the lattice action exactly preserves a subset of the supersymmetry algebra, up to two soft breaking terms that stabilize numerical computations.

\SLAT\ itself uses rational hybrid Monte Carlo importance sampling to compute various observables as discussed in \secref{sec:RHMC}.
The present code allows a number of different $\cN = 4$ SYM studies using larger lattices than could be considered by the serial code presented in \refcite{Catterall:2011cea}, which \SLAT\ supersedes.
In \secref{sec:run} we provided a quick-start guide explaining how to obtain, compile and run this publicly available software~\cite{GitHub}.
Since \SLAT\ is based on the MILC code for lattice QCD, it is relatively easy to modify, extend and improve.
Simulations of $\cN = 4$ SYM are in their infancy, and we encourage researchers to adapt the code as they wish, with the hope of making algorithmic improvements.
To assist such efforts, in \secref{sec:code} we discussed some notable aspects of the documentation, including details of the adjoint representation for arbitrary $N$, data layout, internode communications, MD integrators, and the algorithm we use to compute the Pfaffian in parallel.

While all the applications discussed in this paper are stable and well tested, we are currently extending \SLAT\ to carry out further physics projects, including investigations of scalar correlation functions and of the RG blocking scheme recently proposed by \refcite{Catterall:2014mha}.
Algorithmic improvements are also likely to be explored in the future.
As one example, our MD integrator might be made more efficient by introducing additional time scales to be used by different terms in the partial fraction expansion (\eq{eq:pfrac}), along the lines of the algorithm described in \refcite{open:manual}.
At a purely practical level, we have observed uncomfortably slow thermalization in RHMC runs started from a \verb!fresh! or \verb!random! gauge configuration, especially for larger lattice volumes and larger values of $N$.
Such systems may need to run for more than 2000 MD time units before data taking can begin, introducing a significant overhead cost.
It should be possible to address this issue by designing better initial configurations.

Finally, \SLAT\ should serve as a convenient starting point for parallelized numerical studies of similar lattice theories with extended supersymmetry.
Promising examples include two- and three-dimensional systems (potentially involving matter fields) with lattice actions constructed through topological twisting~\cite{Catterall:2011cea, Joseph:2013jya, Joseph:2013bra, Joseph:2014bwa}.
We have already carried out such an extension for two-dimensional $\cN = (2, 2)$ SYM, which we briefly describe in the appendix.
Much of the parallel framework provided by \SLAT\ should even remain useful for lattice formulations based on different approaches, such as that considered in \refcite{Matsuura:2014pua}.

\vspace{12 pt}
\section*{Acknowledgments} 
We thank our collaborators Simon Catterall, Poul Damgaard and Joel Giedt for many instructive discussions of $\cN = 4$ SYM on and off the lattice, and Aarti Veernala for helping us compare results from \SLAT\ with those from the serial code~\cite{Catterall:2011cea}.
The development of \SLAT\ is supported in part through the Scientific Discovery through Advanced Computing (SciDAC) program funded by the U.S.~Department of Energy (DOE) Office of Science, Office of High Energy Physics under Award Number DE-SC0008669, in addition to support through Award Numbers DE-SC0010005 (TD) and DE-SC0009998 (DS).
Numerical calculations were carried out on the HEP-TH cluster at the University of Colorado and on the DOE-funded USQCD facilities at Fermilab.

\section*{Appendix: Two-dimensional $\cN = (2, 2)$ supersymmetric Yang--Mills theory} 
Although the main focus of \SLAT\ is four-dimensional $\cN = 4$ SYM, we have also adapted the code to consider the similar $\cN = (2, 2)$ SYM theory in two dimensions.
We have implemented this extension as a separate directory tree mirroring that for $\cN = 4$ SYM, rather than guaranteeing that the four-dimensional code will reduce to $\cN = (2, 2)$ SYM upon setting the input parameters \verb!ny = 1! and \verb!nz = 1! (\fig{fig:sample_input}).
This approach is much more computationally efficient, allowing for two dimensions to be dropped completely, at the cost of some code duplication.

The lattice action of two-dimensional $\cN = (2, 2)$ SYM is quite similar to that of $\cN = 4$ SYM:
\begin{align}
  S         & = S_{exact} + S_{stab}                                                                                                                                                                                    \\
  S_{exact} & = \frac{N L N_t}{2\la} \sum_n a^4 \ \Tr{-\cFbar_{ab}(n)\cF_{ab}(n) + \frac{c_2}{2}\left(\cDbar_a^{(-)}\cU_a(n)\right)^2 - \chi_{ab}(n) \cD_{[a}^{(+)}\psi_{b]}^{\ }(n) - \eta(n) \cDbar_a^{(-)}\psi_a(n)} \cr
  S_{stab}  & = \frac{N L N_t}{2\la}\mu^2 \sum_{n,\ c} a^4 \left(\frac{1}{N}\Tr{\cU_c(n) \cUbar_c(n)} - 1\right)^2 + \ka \sum_{\cP} a^4 |\det \cP - 1|^2,                                                               \nonumber
\end{align}
which nearly matches Eqs.~\ref{eq:lattice_action}--\ref{eq:Sstab}.
The main difference is that there are now only two gauge links at each site (forming the familiar square lattice), which forbids the $S_{closed}$ term in \eq{eq:Sclosed}.
The \KD multiplet $\Psi = (\eta, \psi_a, \chi_{ab})$ involves only $2^d = 4$ rather than 16 fermion fields.
Finally, the 't~Hooft coupling \la is now dimensionful, and has to be divided by the lattice volume $L\X N_t$~\cite{Catterall:2011cea}.

The lower dimensionality and reduced field content of $\cN = (2, 2)$ SYM compared to $\cN = 4$ SYM makes the system more tractable numerically.
Parallelization only becomes crucial for lattices larger than $64^2$, for which the overall factor of $\frac{N L N_t}{2\la}$ tends to be rather large.
Since this system has already been fairly well explored using serial codes~\cite{Hanada:2010qg, Catterall:2011cea, Catterall:2011aa, Mehta:2011ud, Galvez:2012sv}, we are not currently carrying out any $\cN = (2, 2)$ SYM physics projects using \SLAT.
Instead, our main interest is using this simpler and cheaper system as a testing ground for observables and algorithms ultimately intended for our $\cN = 4$ investigations.
For example, it was helpful to debug our parallel Pfaffian algorithm (\secref{sec:Pfaffian}) in $\cN = (2, 2)$ SYM before setting up this difficult computation in four dimensions.

Since the \verb!2dSYM! portion of \SLAT\ was adapted from the $\cN = 4$ SYM code discussed in the body of this paper, much of the information presented is directly applicable to this case.
The executables, input parameters and compile-time options are the same as discussed in \secref{sec:run}; the directory structure and most low-level routines are the same as discussed in \secref{sec:code}.
The main differences are the straightforward square lattice and the absence of the $\cQ$-closed term, which greatly simplify the lattice action, forces and internode communications.

\section*{References}
\bibliographystyle{utphys}
\bibliography{par_max_susy}

\newpage
\section*{Program summary}
\setlength{\parindent}{0 mm}
\setlength{\parskip}{6 pt}

{\em Manuscript title:} Parallel software for lattice $\cN = 4$ supersymmetric Yang--Mills theory

{\em Authors:} David Schaich and Thomas DeGrand

{\em Program title:} \SLAT

{\em Journal reference:} 

{\em Catalogue identifier:} 

{\em Licensing provisions:} None 

{\em Programming language:} C

{\em Operating system:} Any, tested on Linux workstations and MPI clusters with InfiniBand

{\em Keywords:} Lattice gauge theory, Supersymmetric Yang--Mills, Monte Carlo methods, Parallel computing

{\em PACS:} 11.15.Ha, 12.60.Jv, 02.70.Uu

{\em Has the code been vectorised or parallelized?:} Code is parallelized

{\em Classification:} 11.5 Lattice Gauge Theory

{\em Nature of problem:} \\
To carry out non-perturbative Monte Carlo importance sampling for maximally supersymmetric Yang--Mills theories in two and four dimensions, and thereby compute observables including Wilson loops, fermion bilinears, eigenvalues of $\cDdag \cD$ and the Pfaffian of the sparse fermion operator $\cD$.

{\em Solution method:} \\
The central application is a rational hybrid Monte Carlo algorithm with a two-level Omelyan molecular dynamics integrator.  Gauge field configurations generated by this application may be saved to disk for subsequent measurements of additional observables.  Input parameters for either configuration generation or analysis may be entered manually or read from a file.

{\em Restrictions:} \\
The code is currently restricted to two-dimensional $\cN = (2, 2)$ and four-dimensional $\cN = 4$ supersymmetric Yang--Mills theories.  The process of topological twisting on which it is based can also be applied to a few other systems, as discussed in Sections~\ref{sec:intro} and \ref{sec:end}.

{\em Additional comments:} \\
Further documentation is provided in the distribution file, including a set of test runs with reference output in the \verb!testsuite! directory.

{\em Running time:} \\
From seconds to hours depending on the computational task, lattice volume, gauge group, and desired statistics, as well as on the computing platform and number of cores used.  For example, rational hybrid Monte Carlo generation of 50 molecular dynamics time units for a \sixt lattice volume with gauge group U(2) takes approximately 16 hours on 512 cores of the USQCD \bc\ cluster at Fermilab, while standard measurements on a saved $8^3\X24$ U(2) configuration require only 8 seconds on one eight-core workstation.

{\em References:} \\
This program supersedes the serial code presented in \refcite{Catterall:2011cea}.

\end{document}